\documentclass[letterpaper,twocolumn,10pt]{article}
\pdfoutput=1
\usepackage[T1]{fontenc}

\usepackage{latexsym,amssymb,stmaryrd}
\usepackage{amsmath}
\usepackage{colortbl}
\usepackage{color}
\usepackage{xspace}
\usepackage{graphics}
\usepackage{graphicx}
\usepackage{multicol}
\usepackage{multirow}
\usepackage{url}
\usepackage{arydshln}
\usepackage{pgfplots}
\usepackage{listings}

\usepackage{enumitem}
\usepackage{ctable}
\usepackage{adjustbox}
\usepackage{hyperref}
\usepackage{tikz}

\usepackage{pgfgantt}

\usepackage{pgfplots}
\usepackage{pgfplotstable}

\usetikzlibrary{
    shapes
  , shadows
  , trees
  , arrows
  , arrows.meta
  , fit
  , calc
  , positioning
  , backgrounds
  , decorations
  , decorations.pathmorphing
  , decorations.text
  , decorations.pathreplacing
  , patterns
  , matrix
  , shadows
  , fit
  , chains
}

\tikzset{onslide/.code args={<#1>#2}{%
  \only<#1>{\pgfkeysalso{#2}} 
}}
\tikzset{temporal/.code args={<#1>#2#3#4}{%
  \temporal<#1>{\pgfkeysalso{#2}}{\pgfkeysalso{#3}}{\pgfkeysalso{#4}} 
}}

\def\bffilldark(#1,#2,#3){
  \foreach \x in {#2,...,#3}{
    \node[draw,fill=gt@gray!30,below=\x em of #1] {};
  }
}

\def\bffilllight(#1,#2,#3){
  \foreach \x in {#2,...,#3}{
    \node[draw,fill=gt@cyan!60,below=\x em of #1] {};
  }
}

\def\bffillred(#1,#2,#3){
  \foreach \x in {#2,...,#3}{
    \node[draw,fill=gt@red!60,below=\x em of #1] {};
  }
}

\tikzstyle{mk} = [draw, fill=gt@red!20, rounded corners,
  minimum height=5.5em, text width=4em, text centered]
\tikzstyle{src} = [draw, fill=gt@yellow!20, rounded corners,
  minimum height=5.5em, text width=4.5em, text centered]
\tikzstyle{exe} = [draw, fill=gt@yellow, rounded corners,
  minimum height=4ex, text width=4em, text centered, text=white]
\tikzstyle{meta} = [draw, fill=gt@blue!80, rounded corners,
  minimum height=4ex, text width=4em, text centered, text=white]
\tikzstyle{var} = [exe, fill=gt@dkred, text=white]
\tikzstyle{db} = [exe, fill=gt@blue, text=white, minimum height=6ex]

\tikzstyle{stage} = [draw, thick, text=white, inner sep=1.5ex]

\tikzstyle{move} = [->,very thick, color=gt@dkgray]
\tikzstyle{connect} = [move,<->]

\tikzstyle{mk} = [draw, fill=gt@red!20, rounded corners,
  minimum height=5.5em, text width=4em, text centered]
\tikzstyle{meta} = [draw, fill=gt@blue!80, rounded corners,
  minimum height=4ex, text width=4em, text centered, text=white]
\tikzstyle{var} = [exe, fill=gt@dkred, text=white]
\tikzstyle{db} = [exe, fill=gt@blue, text=white, minimum height=6ex]
\tikzstyle{database} = [cylinder,
  cylinder uses custom fill,
  cylinder body fill=gt@yellow!30,
  cylinder end fill=gt@yellow!30,
  shape border rotate=90,
  aspect=0.25,
  text width=6em,
  minimum height=0.4em,
  draw ]
\tikzstyle{stage} = [draw, thick, text=white, inner sep=1.5ex]
\tikzstyle{move} = [->,very thick, color=gt@dkgray]
\tikzstyle{connect} = [move,<->]
\tikzstyle{label} = [auto,text=black]
\tikzstyle{gre} = [circle, draw, fill=green!80]
\tikzstyle{blu} = [circle, draw, fill=blue!80]
\tikzstyle{yel} = [circle, draw, fill=yellow]
\tikzstyle{rd}  = [circle, draw, fill=red!80]
\tikzstyle{srcdb} = [draw, fill=gt@yellow!20, rounded corners,
  minimum height=5.5em, text width=4.5em, text centered]
\tikzstyle{exedb} = [draw, fill=gt@yellow, rounded corners,
  minimum height=5.5em, text width=4.5em, text centered, text=white]
\tikzstyle{src} = [draw, fill=gt@yellow!20, rounded corners]
\tikzstyle{exe} = [draw, fill=gt@yellow, rounded corners, text=white]
\tikzstyle{neuron} = [draw, circle, fill=gt@red!20,text width=7ex,text centered]
\tikzstyle{neuron} = [draw, circle, fill=gt@red!20,text width=7ex,text centered]
\tikzstyle{space} = [draw,font=\Large]
\tikzstyle{pop} = [draw, fill=gt@plum!30, circle, text width=6em, text centered]
\tikzstyle{act} = []
\tikzstyle{action} = []
\tikzstyle{vuln} = [src, fill=gt@blue!30, minimum height=3em, text width=7em]
\tikzstyle{bisrc} = [draw, fill=gt@yellow!30, rounded corners, minimum height=4em, text width=5em, text centered]

\tikzstyle{veri} = [move,thick, color=gt@dkred, dashed]
\tikzstyle{base} = [draw, minimum height=4ex, text centered, text width=5em, rounded corners]
\tikzstyle{build} = [base, fill=gt@yellow!20]
\tikzstyle{exe} = [draw, circle, minimum height=3ex, text centered, text width=4em, fill=gt@yellow!20]
\tikzstyle{gt} = [base, fill=gt@red!20]
\tikzstyle{legacy} = [base, fill=gt@dkgray!20]
\tikzstyle{open} = [base, fill=gt@gray!20]
\tikzstyle{move} = [->,very thick, color=gt@dkgray]
\tikzstyle{phase} = [gt@dkred, draw, thick, dotted, inner sep=1em, fill=gt@yellow!10]
\tikzstyle{orig} = [draw, fill=gt@red!20, minimum height=4em, text width=5em, text centered]
\tikzstyle{hard} = [orig, fill=dkgreen!20]
\tikzstyle{cert} = [hard, rounded corners, minimum height=2em]
\tikzstyle{spec} = [draw, fill=gt@red!20, circle, text width=3em, text centered]
\tikzstyle{elicit} = [draw, fill=gt@yellow!30, rounded corners, minimum height=4em, text width=5em, text centered]
\tikzstyle{peephole} = [draw, fill=gt@red!20, text width=8.5em, text centered]
\tikzstyle{neuron} = [draw, circle, fill=gt@red!20,text width=7ex,text centered]
\tikzstyle{space} = [draw,font=\Large]
\tikzstyle{xform} = [draw,rounded corners,fill=gt@red,text=white]
\tikzstyle{tech} = [draw, text centered, text width=5em, rounded corners,font=\small]
\tikzstyle{unit} = [draw,rounded corners,fill=gt@red!20]
\tikzstyle{elicit2} = [draw, fill=gt@yellow!30, rounded corners]
\tikzstyle{bb} = [draw, fill=gt@dkgray!40, text centered, text width=6em, minimum height=3em]
\tikzstyle{bedexe} = [draw, fill=gt@yellow, rounded corners, minimum height=4ex, text width=4em, text centered, text=white]
\tikzstyle{alert} = [text=gt@red]
\tikzstyle{third} = [xform,fill=gt@blue]
\tikzstyle{cicd} = [xform,fill=gt@plum,font=\Huge,ultra thick,inner sep=2em]

\pgfdeclarelayer{bg}    
\pgfsetlayers{background,bg,main}  

\colorlet{dblue}{blue!20!black}

\tikzstyle{data} = [draw,
  text width=6.0em, text centered,
  minimum height=1.5em,
  fill=white,
  text width=4em,
  minimum width=4em,
  minimum height=3em]

\tikzstyle{process} = [draw,
  text width=8.0em,
  text centered,
  fill=blue!10,
  text width=4em,
  minimum width=8em,
  minimum height=3em,
  rounded corners,
  drop shadow]

\tikzstyle{texto} = [above, text width=15em, text centered]
\tikzstyle{line} = [draw, thick, -latex']

\newcommand{\ddisasm}{\ifanonymous\texttt{XXASM}\else\texttt{Ddisasm}\fi\xspace}
\newcommand{\gtirb}{\ifanonymous\textsc{XXIRB}\xspace\else\textsc{GTIRB}\xspace\fi}

\newif\ifanonymous
\anonymousfalse

\newif\ifcasestudies
\casestudiesfalse

\begin{document}

\ifanonymous
\title{\gtirb: Intermediate Representation for Binaries}
\else
\title{\vspace{-3.5em}\gtirb: GrammaTech Intermediate Representation for Binaries}
\author{
  Eric Schulte\quad Jonathan Dorn\quad Antonio Flores-Montoya\quad Aaron Ballman\quad Tom Johnson\\
  GrammaTech, Inc.\\
  \url{{eschulte,jdorn,afloresmontoya,aballman,tjohnson}@grammatech.com}
}
\fi

\maketitle

\begin{abstract}
  \gtirb is an intermediate representation for binary analysis and
  rewriting tools including disassemblers, lifters, analyzers,
  rewriters, and pretty-printers.  \gtirb is designed to enable
  communication between tools in a format that provides the basic
  information necessary for analysis and rewriting while
  making no further assumptions about domain (e.g., malware
  vs. cleanware, or PE vs. ELF) or semantic interpretation (functional
  vs. operational semantics).  This design supports the goals of (1)
  encouraging tool modularization and re-use allowing researchers and
  developers to focus on a single aspect of binary analysis and
  rewriting without committing to any single tool chain and (2) facilitating
  communication and comparison between tools.
\end{abstract}

\section{Introduction}

Software is essential to the functioning of modern societies.
It follows that software analysis, hardening, and rewriting are
essential to the secure and efficient functioning of society.
Unfortunately, software is frequently only available in
binary form: as dependencies of active software projects,
firmware and applications distributed without source, or simply
old software.
Both analysis and rewriting require first lifting
software to an initial intermediate representation (IR).
Binary analysis frameworks typically develop and use their own
internal IR~\cite{ida,ghidra,binja,hopper,bap,radare,b2r2}, in some
cases IRs are borrowed from other tooling such as dynamic analysis
platforms~\cite{angr} or compiler infrastructure~\cite{macaw,pharos}.
In both cases, the IRs typically specify the representation of
instruction semantics, which in turn often dictates the methods of
analysis and the implementation details of their clients.  These
IRs are typically not portable between tools and projects.

\gtirb is intended to facilitate communication between tools for
binary analysis and rewriting.  \gtirb is released as open-source
software\footnote{\ifanonymous
  Withheld.\else\url{https://github.com/GrammaTech/gtirb}\fi} with a
high quality disassembler, \ddisasm,\footnote{\ifanonymous
  Withheld.\else\url{https://github.com/GrammaTech/ddisasm}\fi}
capable of lifting COTS binaries to \gtirb~\ifanonymous
  \cite{anonymous}\else \cite{ddisasm}\fi. To ensure applicability across domains,
  \gtirb's structural requirements are minimal.
To ensure interoperability between tools regardless of their
instruction semantics, \gtirb does not represent instructions; instead
the raw machine-code bytes are stored directly in the IR: they can be
extracted and processed using external tools.
To provide APIs in a variety of programming languages,
\gtirb is serialized using Protobuf~\cite{protobuf}, an efficient
multi-language serialization library.
By enabling communication between binary analysis and rewriting tools
and enabling the modularization of monolithic frameworks, we hope that
\gtirb will promote greater re-use of components across the binary
analysis and rewriting community and lower the barrier of entry for
binary analysis and rewriting research and tool development.

The LLVM project~\cite{lattner2004llvm} demonstrates the huge benefit
a well designed IR can provide to a research community.  
LLVM allows compiler researchers to more easily leverage each other's
work and focus on the problems specific to their own interests.  This
has led to dramatic uptake of LLVM and Clang across both academia and industry.  \gtirb seeks to recreate LLVM's
success for the binary analysis and transformation community.

This paper includes:
\begin{enumerate}[noitemsep,nolistsep]
\item A review of related work in \autoref{related}.
\item An introduction of \gtirb and a high-level description of its
  design in \autoref{design}.
\ifcasestudies
\item A series of case studies to motivate the use of \gtirb for
  binary analysis and rewriting in \autoref{case-studies}.
\fi
\end{enumerate}

\section{Related Work}
\label{related}

Existing binary analysis and rewriting frameworks are frequently
limited by; (i) forcing the choice of instruction semantics (ii)
prescriptive plugin architectures and design assumptions and (iii)
poor support for rewriting.

\paragraph{IDA}
IDA Pro~\cite{idapro} is the industry leading binary analysis and
reverse engineering platform.  It provides disassembly, decompilation,
and an interactive environment for navigating binary programs.  IDA is
extensible through a plugin API, and a sizeable number of
open source plugins have been developed by the community.  The
disassembly produced by IDA is primarily intended to support manual
review and is not intended to support reassembly.

\paragraph{Ghidra}
Ghidra~\cite{ghidra}, recently released by the National Security
Agency (NSA), is a reverse engineering framework that provides an
Eclipse-based graphical user interface.  Like IDA, Ghidra is
extensible, supporting scripts and plugins.  Ghidra is also primarily
intended to support manual analysis and the disassembly and
decompilation provided by Ghidra are not primarily intended for
reassembly or recompilation.  \gtirb supports interoperability with
Ghidra.\footnote{\url{https://github.com/GrammaTech/gtirb-ghidra-plugin}}

\paragraph{Angr}
Angr~\cite{angr} is currently the most widely used binary rewriting
platform.
Angr is distributed via a suite of Python 3
libraries.  The platform provides functionality for disassembly,
analysis, and symbolic execution.
Angr uses the Vex instruction representation from
Valgrind~\cite{nethercote2007valgrind} to represent instructions.
\gtirb boasts faster and more accurate disassembly than Anger's Ramblr
disassembler~\cite{ddisasm}.

\paragraph{BAP}
The CMU Binary Analysis Platform (BAP)~\cite{bap} lifts binaries to
its Binary Intermediate Language (BIL) using tooling based on either
IDA Pro or LLVM.  BIL is only usable through BAPs plugin framework and
does not readily support reassembly.

\paragraph{Uroboros}
Uroboros~\cite{uroboros} was the first tool to focus directly on
generating reassembleable assembly---using a relatively simple
disassembly technique.  Uroboros directly outputs text assembler code.
Rewriting is done by compiling plugins into Uroboros to modify simple
instruction data structures.

\paragraph{Multiverse}
Multiverse~\cite{superset} is a static binary rewriter which does not
use heuristics but reassembles all possible disassemblies.  The
Capstone disassembler is used, and a simple Python API may be used to
add instrumentation.  Only conservative superset-reassembly rewriting
is supported.

\paragraph{LLVM}
LLVM~\cite{lattner2004llvm} provides an IR that is a popular target
for language front-ends, most notably the Clang C/C++ front end.  The
rich ecosystem of optimization and analysis tools implemented over-top
of LLVM make it an attractive target.  There are a number of projects
seeking to lift binary software to LLVM, most notably
McSema~\cite{mcsema} and SecondWrite~\cite{secondwrite}.
Unfortunately LLVM is a difficult target for binary lifting given the
strongly typed memory model, which forces very difficult analysis
decisions before the IR may even be constructed.


\paragraph{Debug File Formats} Established debug file formats such as PDB (Program
Database)~\cite{mspdb}, DWARF~\cite{dwarf}, and stabs~\cite{stabs}
typically provide type, symbol, and location information designed for
mapping elements of a binary onto original source code to aid in
debugging and related activities.\footnote{
  \begin{minipage}{0.425\textwidth}
    We checked StackOverflow~\cite{stackoverflow} for all questions
    tagged with one or more of these technologies as of 3/19/2020. Only
    a tiny proportion could be identified as referring to use of the
    format without explicit mention of debugging/profiling, and none of those gave
    any specific indication of an alternative use in practice.
    \begin{tabular}{|l|l|l|l|}
      \hline
      Format & Tags           & Total & Not explicitly  \\
             & Posts          &       & Debugging posts \\
      \hline
      PDB    & \tt{pdb-files} & 410   & 2               \\ 
      DWARF  & \tt{dwarf}     & 216   & 0               \\ 
      stabs  & \em{no tag}    & -     & -               \\ 
      \hline
    \end{tabular}  
  \end{minipage}
} Various common characteristics of these formats render them
unsuitable as an intermediate representation for reassembly or
rewriting: the formats are often proprietary, incompletely documented,
or both; they must be generated with full source code and compilation
details in order to be properly populated; they do not preserve all
binary information but must instead be used in concert with the
original binary (and often the original source as well).

\section{Design of \gtirb}
\label{design}

\begin{figure*}
  \centering
  \adjustbox{max size={\linewidth}{\textheight}}{
\tikzstyle{object} = [draw]
\tikzstyle{owner} = [thick]
\tikzstyle{owner-many} = [-triangle 90 reversed,thick]
\tikzstyle{reference} = [owner, dashed]
\begin{tikzpicture}

  \node[object, ] (ir) {IR (1)};
  \node[object, below right=1em and 2em of ir, text width=6.25em] (auxdata) {AuxData\\
    \vspace*{0.25em}
    \hspace*{-1em}
    \begin{tabular}{r|l}
      ID1 & DATA1 \\
      ID2 & DATA2 \\
      ID3 & DATA3 \\
      ID4 & DATA4
    \end{tabular}
  };
  \node[right=2em of auxdata] (anything) {{\em any UUID}};
  \node[object, above right=4em and 2em of ir] (cfg) {IPCFG (1)};
  \node[object, right=4em of cfg] (edges) {Edges};

  \node[object, right=4em of ir] (modules) {Modules};
  \node[object, below right=1em and 4em of modules] (symbols) {Symbols};

  \node[object, right=4em of modules] (sections) {Sections};

  \node[object, right=2em of sections] (byteintervals) {ByteIntervals};

  \node[object, above=2em of byteintervals] (proxies) {Proxy blocks};
  \node[object, right=2.5em of proxies] (codeblocks) {Code Blocks};
  \node[object, below=0.5em of codeblocks] (datablocks) {Data Blocks};
  \node[fit={(proxies) (codeblocks)}, draw, thick, dotted] (edgeblocks) {};
  \node[above=0em of edgeblocks.165] (eblocks) {Edge Blocks};
  \node[fit={(codeblocks) (datablocks)}, draw, thick, dotted, inner sep=0.75em] (byteblocks) {};
  \node[below=0em of byteblocks] (bblocks) {Byte Blocks};

  \node[object, below right=1em and 2.5em of byteintervals] (symexpr) {SymbolicExpressions};

  \draw[owner-many] (ir) -- (modules);
  \draw[owner] (ir) -- ++(1.0,0) |- (cfg);
  \draw[owner-many] (ir) -- ++(1.0,0) |- (auxdata);
  \draw[owner-many] (cfg) -- (edges);
  \draw[owner-many] (modules) -- ++(1.5,0) |- (symbols);
  \draw[owner-many] (modules) -- (sections);
  \draw[owner-many] (modules) -- ++(1.5,0) |- (proxies);
  \draw[owner-many] (modules.south) -- (auxdata.north-|modules.south);
  \draw[owner-many] (sections) -- (byteintervals);
  \draw[owner-many] (byteintervals) -- ++(1.75,0) |- (symexpr);
  \draw[owner-many] (byteintervals) -- ++(1.75,0) |- (bblocks);

  \draw[reference] (edges.east) |- (eblocks);
  \draw[reference] (symexpr.south) |- ++(-4,-1) |- (symbols.east);
  \draw[reference] (auxdata) -- (anything);

\end{tikzpicture}

}
  \caption{\label{structure}\gtirb data structure.  Solid lines denote
    inclusion/ownership.  Dashed lines denote references by UUID.  A
    $\blacktriangleleft$ ending of an arrow denotes a one-to-many
    relationship where the owning object may hold any number of the
    owned element.}
\end{figure*}
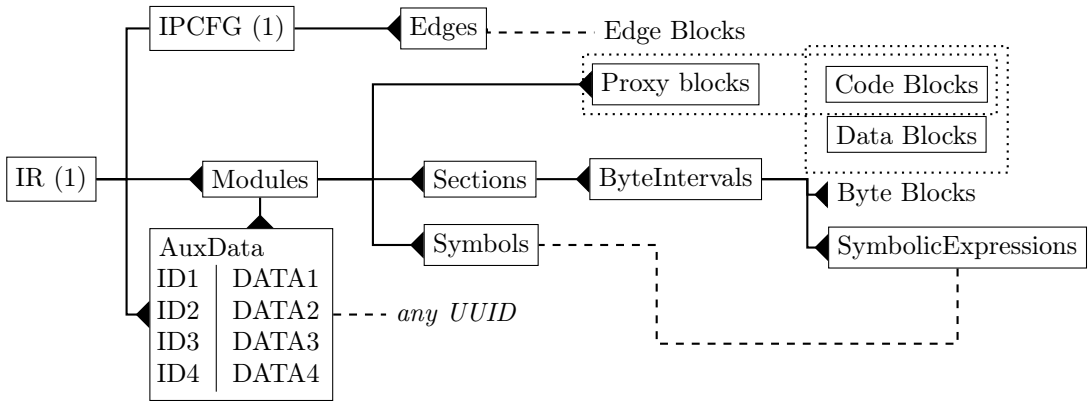

An instance of \gtirb is a single data structure organized as shown in
\autoref{structure}.

\subsection{Core Structures}
\label{core-structures}
At the top level of every \gtirb instance is a single {\em IR}
element.  This IR holds multiple {\em Modules}.  Each Module
corresponds to a single compilation unit, such as an executable or a
shared library. For example, a single \gtirb IR could represent a
binary executable and all of the libraries it uses dynamically, with
each library represented as a separate {\em Module}.

The primary contents of each Module are the {\em CodeBlocks} and the
{\em DataBlocks}, which represent the Module code and data
respectively; each block belongs to a {\em Section} of the Module.
Both CodeBlocks and the DataBlocks store their contents as regions of
raw bytes in one of the {\em ByteIntervals} associated with their
Section.  The ByteIntervals are vectors of bytes holding the raw
contents of the module.  The ranges of both CodeBlocks and DataBlocks
may overlap arbitrarily with the ranges of other CodeBlocks and
DataBlocks in the same ByteInterval.  All bytes in the module are
stored in ByteIntervals.  The granularity of the ByteIntervals (many
small ByteIntervals vs. fewer larger ByteIntervals) determines the
flexibility of rewriting supported by a GTIRB instance.

\gtirb does not explicitly indicate the interpretation of the bytes in
CodeBlocks or DataBlocks.  CodeBlocks are notionally intended to
represent basic blocks of instructions, though this is not enforced.  A
separate decoder is required to extract individual instructions from a
code block (see \autoref{instruction-storage}).
Similarly \gtirb forces no interpretation of a DataBlock's bytes.
Any deduced type for a DataBlock may be stored as {\em auxiliary data}
(see \autoref{aux-data-tables}).

\gtirb imposes an additional level of structure on code.  The {\em
IPCFG} is a single graph covering all code in the IR (see
\autoref{cfg-or-ipcfg}) in which each node is a block and each edge
connects two blocks (see \autoref{byte-types-and-edge-types}).
Edges between CodeBlocks represent control flow in the IPCFG.
Special {\em ProxyBlocks} are used to represent control flow sources
or targets that cannot be resolved to CodeBlocks---e.g., control flow
to libraries not included in the IR instance.

\gtirb explicitly represents {\em Symbols} and {\em
  SymbolicExpressions}.  These provide symbolization information for
CodeBlocks and DataBlocks.  For CodeBlocks, they indicate which
operands are symbolic to specific instructions; for DataBlocks, which
data is symbolic.
SymbolicExpressions are associated with offsets in their containing
ByteInterval.  This places them at particular locations in the
CodeBlocks or DataBlocks (i.e., {\em ByteBlocks} in
\autoref{structure}) held in the ByteInterval.  
The symbolization information provided by \gtirb is sufficient to
enable the contents of the binary to be reorganized in memory while
maintaining all cross references to preserve binary functionality.

Every element of \gtirb, namely: {\em Modules}, {\em Symbols}, {\em
  SymbolicExpressions}, {\em Sections}, {\em ByteIntervals}, {\em
  DataBlocks}, {\em CodeBlocks}, and {\em Edges} has a universally
unique identifier (UUID).
UUIDs
allow both first-class IR components and AuxData tables to reference
other elements of the IR in a manner that is robust to rewriting.
For example, {\em Edges} use UUIDs to reference blocks.
Note that reference by
address in the original binary would not be robust to rewriting as new
entities could not be added to the IR without synthesizing {\em fake}
addresses.

\gtirb is serialized using Google's protobuf~\cite{protobuf},
making it possible to efficiently read and write \gtirb from any
language with Protobuf support.
Currently there are mature \gtirb libraries in C++, Python, and Common
Lisp: these provide more ergonomic and efficient APIs than the default
Protobuf APIs.

\subsection{Auxiliary Data Tables}
\label{aux-data-tables}
The core \gtirb data structure described in \autoref{core-structures}
is intentionally very sparse.
Even very generally useful information, such as the concept of
functions, is not included by default because its use may not be
universal, e.g. malware or hand-written assembler may not have
functions.  One of the core purposes of \gtirb is to communicate
analysis results between tools but the only analyses explicitly
representable in the core \gtirb structure are symbolization, CFG, and
code vs. data.  Much of the information of any instance of \gtirb is
intended to be communicated not in the core required structures, but
instead via auxiliary data ({\em AuxData}) tables.  These tables are
extensible and may be used to store maps and vectors of basic \gtirb
types or arbitrary data in a portable way.  AuxData tables make heavy
use of UUIDs to reference elements of the core \gtirb IR.

\gtirb specifies a small number of ``sanctioned'' AuxData table
schemata to support tool interoperability. These cover very common
information requirements, such as function boundaries.  Current
sanctioned schemas are listed in \autoref{sanctioned-tables}: we
anticipate extending this set as \gtirb becomes more widely used and
new common use cases emerge.

\begin{table*}
  \centering
  \caption{\label{sanctioned-tables}Sanctioned AuxData table schemas,
    expressed as C++ types.}
  \begin{tabular}{l|l}
    Label            & Type \\
    \hline
    {\tt functionBlocks}   & \verb|std::map<gtirb::UUID, std::set<gtirb::UUID>>| \\
    {\tt functionEntries}  & \verb|std::map<gtirb::UUID, std::set<gtirb::UUID>>| \\
    {\tt functionNames}    & \verb|std::map<gtirb::UUID, gtirb::UUID>| \\
    {\tt types}            & \verb|std::map<gtirb::UUID, std::string>| \\
    {\tt alignment}        & \verb|std::map<gtirb::UUID, uint64_t>| \\
    {\tt comments}         & \verb|std::map<gtirb::Offset, std::string>| \\
    {\tt symbolForwarding} & \verb|std::map<gtirb::Symbol,gtirb::Symbol>| \\
  \end{tabular}
\end{table*}

The sanctioned tables in \autoref{sanctioned-tables} have the following
meanings.
\begin{description}
\item[{\tt functionBlocks}] This table identifies function boundaries.
  A function is stored as a set of code blocks.  Storage as a set
  instead of a region of memory ensures robustness to modification of
  the IR and permits the representation of non-contiguous functions.
\item[{\tt functionEntries}] Stores the set of CodeBlocks used as entry
  points to a function.
  Representation of multiple-entry functions is
  supported.
\item[{\tt functionNames}] Stores a single name (as a Symbol
  reference) for each function.  Note that a UUID for a function is
  generated specifically in order to identify the function within and
  across the {\tt functionBlocks}, {\tt functionEntries}, and {\tt
    functionNames} AuxData tables.
\item[{\tt types}] Records the type of a DataBlock, 
  as a string containing a valid C++ type specifier.
\item[{\tt alignment}] Indicates the preferred alignment of a
  CodeBlock or DataBlock in memory (see \autoref{padding}).
\item[{\tt comments}] Associates arbitrary comment
  strings with specific offsets within blocks
  (for examples, an instruction in a CodeBlock or a particular point within
  a DataBlock).
\item[{\tt symbolForwarding}] Redirects one Symbol to another. This is
  useful to resolve indirections related to dynamic linking. For
  example, in ELF files it connects Symbols pointing to Procedure
  Linkage Table (PLT) entries to the function Symbols called in such
  entries.  It also resolves indirect references via the Global Offset
  Table (GOT) used to resolve addresses of global variables at
  runtime.
\item[{\tt padding}] Records inserted padding as a (location,length) pair.
\end{description}

\subsection{Design Decisions}
\label{design-decisions}
Many decisions were made in the design of \gtirb.  These were
motivated by (i) decades of; experience in the development and use of
tools for binary analysis and rewriting, (ii) a desire to maximize
generality and flexibility, and (iii) a desire for simplicity and
orthogonal elementary concepts when possible.  In this section we
discuss some of the potentially more surprising decisions we made.

\subsubsection{Instruction Storage}
\label{instruction-storage}

The most frequent misconception about \gtirb is that it is an
intermediate language (IL) for representing the {\em semantics} of
assembler instructions in the same way that BAP's
BIL,\footnote{\url{https://github.com/BinaryAnalysisPlatform/bil/releases/download/v0.1/bil.pdf}}
Angr's Vex,\footnote{\url{{https://github.com/angr/pyvex}}} or
Ghidra's P-code are ILs.
\gtirb represents the higher-level structure of the binary while preserving the {\em content} of assembler instructions for use by client applications.
The
structural elements are often the result of sophisticated analyses such as
those performed by our \ddisasm front end in recovering the various IR
components: IPCFG (indirect jump target discovery, disassembly);
Symbols and Symbolic Expressions (symbolization); CodeBlocks (code
localization, disambiguation); DataBlocks and {\tt types} AuxData
table (data location, typing); function-related AuxData tables
{\tt functionBlocks}, {\tt functionEntries}, {\tt functionNames}.

\gtirb uses the most general and efficient instruction representation
available---possibly some of the most over-engineered serialization
encodings in human history---the raw machine code bytes.  The users of
\gtirb may read/write these bytes using the decoder/encoder of their
preferred IL (e.g., BIL, Vex, P-code) or using the high quality
open-source
Capstone\footnote{\url{https://www.capstone-engine.org}}/Keystone\footnote{\url{{https://www.keystone-engine.org/}}}
libraries.

This decision has the benefits of universality and memory efficiency.
Abstract syntax tree (AST) representations of instructions often incur very large space
overheads of many times the space required by the machine code bytes.
This overhead is often {\em the} limiting factor when analyzing large
binaries, or collections of models.  Decoding and encoding
machine-code bytes as needed permits fast access to an {\em extremely}
efficient representation of the code.  The universality of
machine-code bytes ensures that the core mission of interoperability
is not compromised, and even permits useful flexibility within a
single project or framework.

The main drawback to this decision is that \gtirb does not provide
instruction semantics.  However, there are already many powerful tools
in this space, such as those referenced at the beginning of this
section, as well as emerging standards.  In our experience using
\gtirb with our own custom instruction semantics and with
Capstone/Keystone,\footnote{\url{https://github.com/GrammaTech/gtirb-capstone}}
the access patterns required by machine code bytes are manageable and
well worth the benefits.

\subsubsection{CFG vs. IPCFG}
\label{cfg-or-ipcfg}
The use of an IPCFG instead of a typical CFG with functions is a
result of the choice to not have first class functions
(\autoref{second-class-functions}).
Dispensing with the intermediate decomposition of the CFG into
functions (i.e. procedures) simplifies both construction and use of
the IPCFG in many cases.
A significant advantage of this approach is that subsequent
analyses only depend on detangling the
often tricky edge cases of function boundary identification when those
analyses explicitly require this information.  Forcing the encoding of
functions into a CFG would make this an implicit potential source of
error for any analysis using the CFG, even those which don't require
function information.

The IPCFG also opens the door to non-standard code representations,
such as dispensing with the notion of basic blocks and instead
representing the code section as a graph of single instructions joined
by control flow edges---as done by SEI's Pharos~\cite{pharos}.  (This
is easily represented in \gtirb using single-instruction code blocks.)

\subsubsection{Second-class functions}
\label{second-class-functions}

Functions are not essential to a functioning binary. For example, malware and
hand-written assembler may dispense entirely with the function abstraction.
Even in compiled code, function boundary identification is a difficult
problem and an active research
area~\cite{meng2016binary,byteweight,shin2015recognizing}. However,
many static analyses and transformations require function boundaries
to work. Thus, we allow for the representation of functions as sets of
basic blocks, entry points and names in AuxData tables---and in a
supporting library.\footnote{\url{https://github.com/GrammaTech/gtirb-functions}}  This also
simplifies the CFG representation.

\subsubsection{Block types and Edge types}
\label{byte-types-and-edge-types}
CodeBlocks represent a range of bytes within a ByteInterval in a
single Section that
are interpreted as code. (Bytes interpreted as data are represented by
DataBlocks.) The bytes covered by each CodeBlock may include
a number of distinct instructions.
Although \gtirb does not represent these individual instructions
explicitly, the conventional use-case is that each CodeBlock contains
only a single basic block so that instruction decoding is trivial and
unambiguous. That is, non-local control flow such as branches, calls,
and returns occurs only at the end of a Block.  However, there is no
enforcement of this convention and in some cases it may be useful to
represent an entire {\tt .text} section as a single code block.
At the other end of this spectrum, single-instruction CodeBlocks and
the corresponding IPCFG are also comfortably represented in \gtirb.

CodeBlocks and ProxyBlocks constitute the nodes of the \gtirb IPCFG;
the control flow between Blocks is encoded as labeled edges.
Edge labels incorporate multiple dimensions:
conditional/unconditional, direct/indirect, and control flow type
(fallthrough, branch, call, return, system call, system call return,
etc.)
These dimensions on edge labels result in an expressive IPCFG.  

\subsubsection{Extra-IR edges}
\label{extra-module-edges}

To represent control flow between CodeBlocks in different IRs, \gtirb
uses ProxyBlocks.  A ProxyBlock may be used as a node in the
IPCFG---as either the source or the target of an edge---but has no
corresponding range of bytes.

For example, to represent a call to a function defined outside of the
current IR, a client may insert a proxy into the IPCFG to represent the
external function, then insert an edge between the calling block and
the proxy.  Similarly, if desired, a call from an external block can
be represented by introducing a proxy to represent the caller and an
edge from that proxy to the entry block of the called function.

\subsubsection{Explicit Padding vs. Alignment}
\label{padding}
Compiler-generated padding between functions in the code section of a
binary could be represented as ``code'' (i.e. {\tt nop}s),
``padding,'' or ``None'' simply not represented.  \gtirb takes the
``None'' option by adding alignment directives to CodeBlocks (see
``{\tt alignment}'' in \autoref{sanctioned-tables}) instead of
explicitly representing code or padding.  This avoids both the space overhead
of the 'Code' option and the increased representation and processing
complexity of the 'Padding' option.

\section{Conclusion}

\gtirb is an intermediate representation of the structure of binaries,
intended to facilitate communication between tools for binary analysis
and rewriting.  An explicit design goal has been to maximize
flexibility and extensibility while providing a minimal core
structure.
This supports incremental lifting and analysis, since
additional analysis results may be added in the form of refined \gtirb
structures or new AuxData tables in subsequent phases.
\gtirb's language-agnostic serialized format also encourages
interoperation between tools written in many languages and
on top of different analysis frameworks and semantics, through the
medium of a language-agnostic serialization format.
We hope that open-sourcing \gtirb and our high-quality \ddisasm
frontend will stimulate a robust ecosystem of interoperable binary
analysis and rewriting tools.

\section{Acknowledgments}
Many thanks to our colleagues at GrammaTech who contributed to the
design and implementation of \gtirb especially; Brian Alliet, Abhishek
Bhaskar, John Farrier, Amy Gale, and Nathan Weston.

This material is based upon work supported by the Office of Naval
Research under contract No. N68335-17-C-0700. Any opinions, findings
and conclusions or recommendations expressed in this material are
those of the authors and do not necessarily reflect the views of the
Office of Naval Research.

\bibliographystyle{plain}
\bibliography{bib/bibliography}

\end{document}